\documentclass[graybox]{svmult}

\usepackage{mathptmx}       
\usepackage{helvet}         
\usepackage{courier}        
\usepackage{type1cm}        
\usepackage{makeidx}         
\usepackage{graphicx}        
\usepackage{multicol}        
\usepackage[bottom]{footmisc}

\usepackage{epstopdf}

\makeindex             

\begin{document}

\title*{Studying the hybrid pulsator 12 Lacertae: mode identification and complex seismic modelling }
\author{Szewczuk Wojciech, Walczak Przemys{\l}aw and Daszy\'nska-Daszkiewicz Jadwiga}
\institute{Szewczuk W., Walczak P., Daszy\'nska-Daszkiewicz J. \at Instytut Astronomiczny, Uniwersytet Wroc{\l}awski,
Kopernika 11, 51-622 Wroc{\l}aw, Poland, \email{szewczuk@astro.uni.wroc.pl, walczak@astro.uni.wroc.pl, daszynska@astro.uni.wroc.pl}}
%
%
\maketitle

\abstract*{We present identification of the mode degree, $\ell$, for all observed frequencies of 12 Lac and results of seismic modelling
which consists in fitting simultaneously the centroid mode frequencies and the corresponding values of the complex nonadiabatic
$f$-parameter. Effects of chemical composition, opacities, core overshooting
and non-LTE atmospheres were taken into account.}

\abstract{We present identification of the mode degree, $\ell$, for all observed frequencies of 12 Lac and results of seismic modelling
which consists in fitting simultaneously the centroid mode frequencies and the corresponding values of the complex nonadiabatic
$f$-parameter. Effects of chemical composition, opacities, core overshooting
and non-LTE atmospheres were taken into account.}

\section{Mode identification}
\label{sec:1}
12 Lac is the early B-type pulsator in which at least ten frequencies of the $\beta$ Cep type and one of the SPB type
are excited.
To determine the pulsation mode degree, $\ell$, we made use of the amplitudes and phases
in the Str\"omgren $uvy$ photometric passbands \cite{handler} as well as the radial velocity data \cite{desmet}.
Two approaches were used: the first one with the theoretical $f$-parameter \cite{jdd2002} and the second
one with the empirical values of $f$ \cite{Daszynska}. The LTE \cite{Kurucz} and non-LTE \cite{Lanz}
stellar atmosphere models were included.
We were able to find identification for five strongest modes
and limitation on $\ell$ for the others.
The dominant mode $\nu_1$ as well as $\nu_2$ are certainly the dipole modes.
Identification of $\nu_3$ and $\nu_5$ is also unique, they are quadruple modes,
and $\nu_4$ is a radial mode. The high-order g-mode is the dipole.

\section{Seismic models of 12 Lac}
\label{sec:2}

\begin{figure}[t]
\sidecaption
\includegraphics[clip=true, scale=.32]{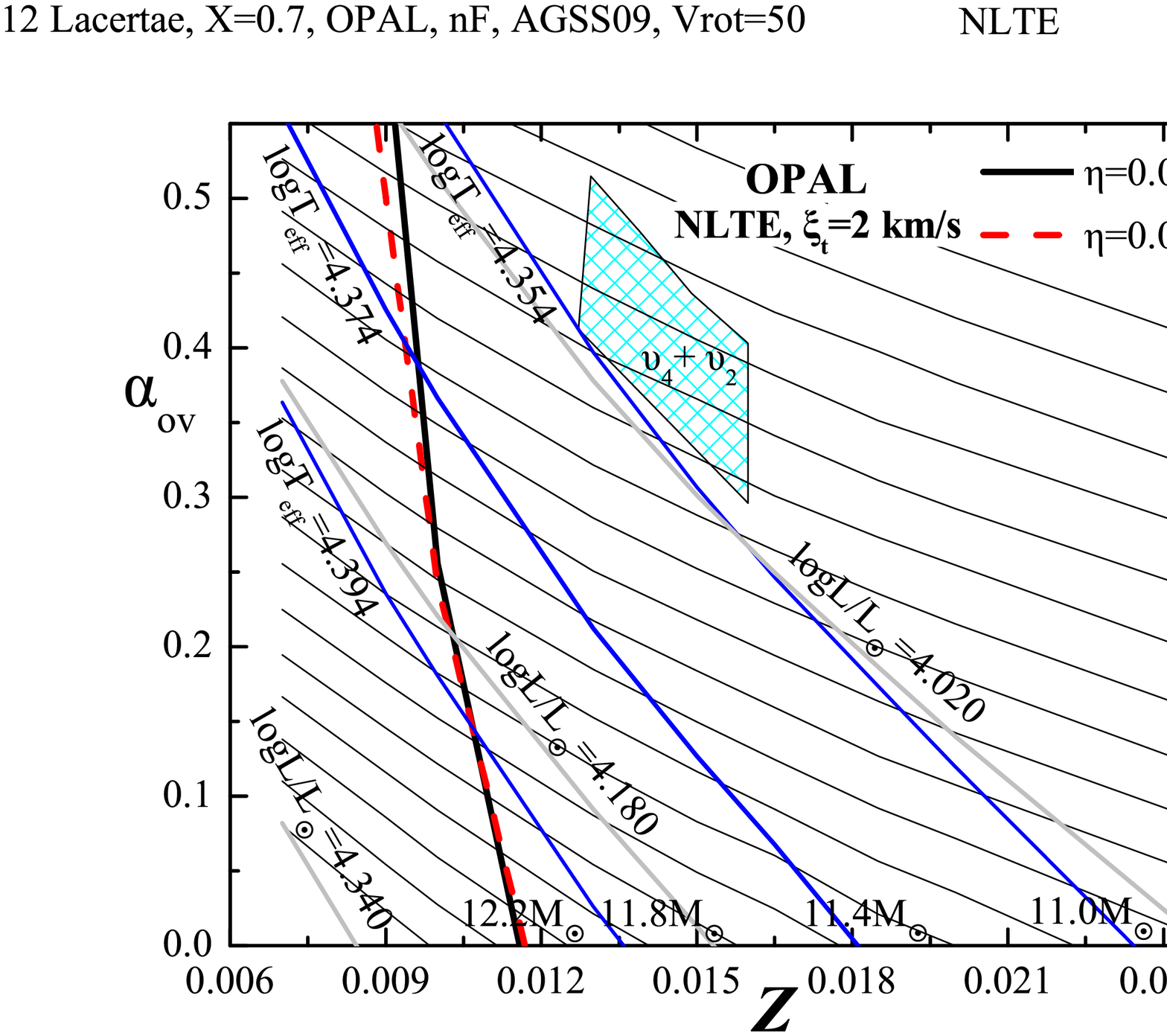}
%
%
\caption{The OPAL seismic models of 12 Lac fitting $\nu_2$ and $\nu_4$ on the $Z-\alpha_{ov}$ plane.
The hatched area indicates models fitting the empirical $f-$parameter, simultaneously for $\nu_4$  and $\nu_2$.
The empirical values of $f$ were calculated with the NLTE atmospheres.}
\label{fig:1}       
\end{figure}

We managed to construct models fitting simultaneously the two frequencies, $\nu_4$ (the radial p$_1$ mode)
and $\nu_2$ (the dipole g$_1$ mode),
and the corresponding values of the $f$-parameters. This approach is called complex asteorseismology
and yields more information on model and theory.
Calculations were performed in a wide range of metallicity,
overshooting parameter from the convective core as well as for different stellar atmosphere models:
LTE with the microturbulent velocity of 2 and 8 km/s, and NLTE with 2 km/s.
We used the OPAL and OP opacity data, and the AGSS09 chemical mixture \cite{Asplund09}.
All effects of rotation were ignored.
In Fig.\,\ref{fig:1}, we give an example of seismic models obtained with the OPAL data on the $Z-\alpha_{ov}$ plane.

%

\section{Conclusions }
Mode identification with the LTE and non-LTE atmospheres provided similar results.
Comparison of the empirical and theoretical values of the nonadiabatic $f$-parameter allowed us
to determine the radial order of the radial mode $\nu_4$. Clearly, $\nu_4$ is the fundamental mode.
With both opacity data we were able to construct seismic models fitting
the $\nu_4$ and $\nu_2$ frequencies and the corresponding $f$-parameters. Moreover, models fitting the high-order
g-mode frequency and reproducing its value of $f$ were found. No preferences for any opacity tables were obtained.

%

%

\end{document}